\documentclass[cameraready]{my_style}

\title{StyleStream: Real-Time Zero-Shot Voice Style Conversion}

\author[affiliation={1}]{Yisi}{Liu}
\author[affiliation={1}]{Nicholas}{Lee}
\author[affiliation={1}]{Gopala}{Anumanchipalli}

\address{
    $^1$ UC Berkeley, USA 
}

\email{louis\_liu@berkeley.edu}


\usepackage{comment}

\begin{document}

\maketitle

\begin{abstract}
    Voice style conversion aims to transform an input utterance to match a target speaker’s timbre, accent, and emotion, with a central challenge being the disentanglement of linguistic content from style. While prior work has explored this problem, conversion quality remains limited, and real-time voice style conversion has not been addressed. We propose StyleStream, the first streamable zero-shot voice style conversion system that achieves state-of-the-art performance. StyleStream consists of two components: a Destylizer, which removes style attributes while preserving linguistic content, and a Stylizer, a diffusion transformer (DiT) that reintroduces target style conditioned on reference speech. Robust content-style disentanglement is enforced through text supervision and a highly constrained information bottleneck. This design enables a fully non-autoregressive architecture, achieving real-time voice style conversion with an end-to-end latency of 1 second. 
    Samples and real-time demo: \url{https://berkeley-speech-group.github.io/StyleStream/}.
\end{abstract}

\section{Introduction}
\begin{figure*}[t]
    \centering
    \includegraphics[width=\linewidth]{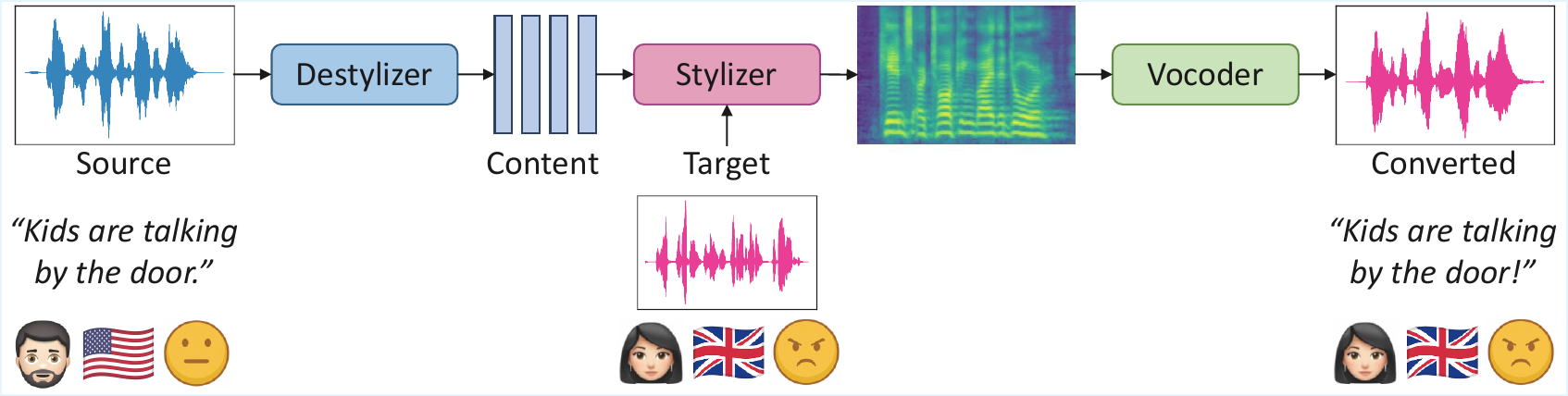}
    \caption{System overview of StyleStream. The Destylizer extracts content features disentangled from style, and the Stylizer generates speech that preserves the source linguistic content while adopting the target timbre, accent, and emotion.}
    \label{system}
\end{figure*}
\begin{figure*}[t]
    \centering
    \includegraphics[width=\linewidth]{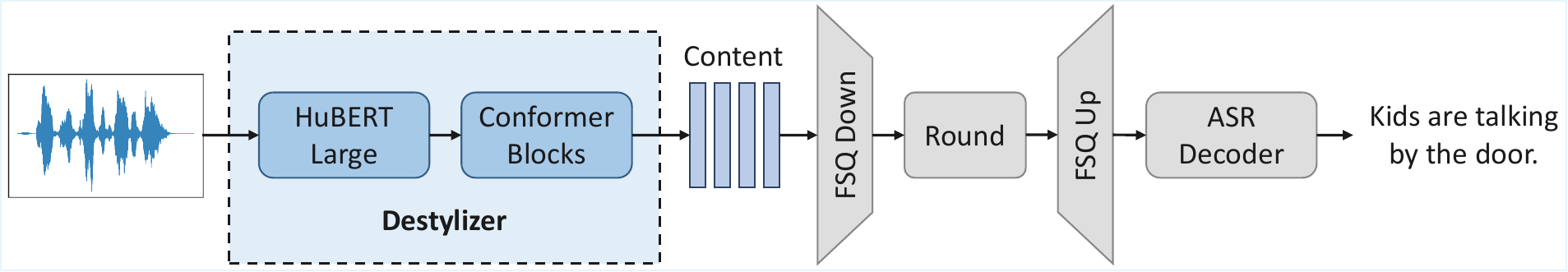}
    \caption{Destylizer architecture. The Destylizer, as part of the ASR encoder, is trained with a sequence-to-sequence ASR loss. The continuous representations immediately before the FSQ module are taken as content features.}
    \label{Destylizer}
\end{figure*}

Zero-shot voice style conversion seeks to modify an input utterance so that it reflects the timbre, accent, and emotion of an unseen target speaker (collectively defined in this work as \textit{\textbf{voice style}}) while preserving the original linguistic content, using only a few seconds of reference speech. Although zero-shot voice style cloning has been extensively studied in the text-to-speech (TTS) domain \cite{voicebox, e2tts, f5tts, maskgct, seedtts, cosyvoice, cosyvoice2}, progress on zero-shot voice style conversion remains limited. In TTS, the linguistic content is taken directly from text, which is fully disentangled from the target speech that provides voice style. In contrast, zero-shot voice style conversion requires the model to first extract the content features from the source utterance, and then generate speech in the target style. This content extraction relies on clean content-style disentanglement, which is the core challenge in speech-to-speech conversion.

Existing strategies for disentanglement include information bottleneck \cite{autovc, triple-bottleneck, naturalspeech3, cosyvoice, cosyvoice2, vevo}, signal perturbation \cite{nansy, nansy++, contentvec, speechsplit2}, mutual information minimization \cite{vqmivc, Yu-disentanglement}, etc. While these approaches achieve reasonable content-timbre disentanglement, the extracted features often remain entangled with accent and emotion. For example, CosyVoice 2 \cite{cosyvoice2} combines information bottlenecks with ASR for supervised disentanglement, yet its “semantic tokens” still encode considerable accent and emotion information due to the large codebook size (6561), as demonstrated in Section \ref{sec:content_feature_ablations}. The current state-of-the-art framework, Vevo \cite{vevo}, employs VQ-VAE \cite{vqvae} to quantize HuBERT \cite{hubert} features with a compact codebook of size 32, producing discrete content tokens. Although Vevo demonstrates good conversion performance, its purely self-supervised training provides no guarantee on what information is discarded by the bottleneck, and linguistic content is prone to degradation during quantization, leading to suboptimal intelligibility and conversion quality (see Section \ref{exp}). Beyond disentanglement, another open challenge is real-time application: while real-time voice conversion (timbre only) has been widely studied \cite{rt-vc, streamvc, streamvoice}, there is no existing work that addresses real-time voice style conversion (timbre, accent and emotion), leaving an important gap.

Motivated by these challenges, we present StyleStream, a novel framework that, for the first time, supports zero-shot real-time voice style conversion across timbre, accent, and emotion, achieving state-of-the-art performance. As shown in Figure \ref{system}, the Destylizer first extracts content features that are disentangled from style information. The Stylizer then takes these content features and, conditioned on the target speech, generates the stylized mel-spectrogram, which is subsequently converted into waveform by the vocoder. 

To achieve better content-style disentanglement, inspired by CosyVoice 2 \cite{cosyvoice2}, we train the Destylizer with an ASR loss and apply finite scalar quantization (FSQ) \cite{fsq} as an information bottleneck, but unlike CosyVoice 2, where a large codebook size (6561) is used, we constrain the codebook size to 45. This combination of text supervision and a compact codebook enables cleaner disentanglement. The choice of content features is also crucial: rather than using the discrete codes directly, we adopt the continuous representations immediately preceding the FSQ layer as content features, following the intuition of SoftVC \cite{softvc}. As shown in Section \ref{ablation}, this design is another critical factor for effective disentanglement. 

For the Stylizer, we train a diffusion transformer (DiT) \cite{dit} with a spectrogram inpainting objective, similar to \cite{voicebox, e2tts, f5tts}. Since no autoregressive modules are involved, the input and output lengths remain identical, which makes streaming straightforward and avoids lag or overlap caused by input-output length mismatches.

In summary, our contributions include:
\begin{itemize}
    \item We introduce StyleStream, the first real-time voice style conversion system with an end-to-end latency of 1 second.
    \item By combining ASR loss and a compact quantization codebook, the Destylizer achieves cleaner content-style disentanglement compared to existing methods.
    \item Trained on 50k hours of English data, StyleStream delivers state-of-the-art conversion quality, substantially improving accent and emotion similarity over prior work.
\end{itemize}

\section{Related Work}
\subsection{Voice Style Cloning}

Voice style cloning aims to reproduce a target speaker’s timbre, accent, and emotion from reference speech. A large body of prior work has explored this problem in the text-to-speech (TTS) domain, where textual input provides a clean and naturally disentangled representation of linguistic content. Modern TTS-based voice style cloning approaches can be broadly categorized into two architectural classes: (1) non-autoregressive models, which primarily rely on diffusion transformers trained with feature inpainting objectives~\cite{voicebox, e2tts, f5tts, maskgct}; and (2) autoregressive models, which condition on text as a prefix and generate speech tokens using next-token prediction~\cite{valle, seedtts, cosyvoice, cosyvoice2, indextts2}. These methods have demonstrated impressive synthesis quality and style fidelity, but their reliance on text limits their applicability to speech-to-speech conversion scenarios.

In contrast, speech-to-speech voice style conversion operates directly on acoustic signals and thus requires explicit content–style disentanglement. Most prior work in this area focuses on transferring individual style attributes in isolation, such as timbre \cite{autovc, stargan-vc, seedvc, hierspeech++}, accent \cite{convspeak, accent_2, accent_3}, or emotion \cite{emovox, emotion_2, emotion_3}. While effective for targeted attribute transfer, these methods do not enable holistic voice style cloning that jointly matches the target speaker’s timbre, accent, and emotion.

Only a limited number of works attempt to jointly clone timbre, accent, and emotion in a unified speech-to-speech setting. Among existing approaches, Vevo~\cite{vevo} is, to our knowledge, the only system that explicitly targets holistic voice style cloning. However, Vevo operates in an offline setting, relies on non-streamable architectures, and exhibits a clear tradeoff between content preservation and style fidelity. These limitations highlight the difficulty of achieving clean content-style disentanglement while maintaining high-quality and streamable speech generation.

In this work, we address these challenges by proposing a streamable speech-to-speech framework that performs holistic voice style cloning, jointly transferring timbre, accent, and emotion from reference speech while preserving high intelligibility.

\subsection{Speech Content Disentanglement}
A key difficulty in speech-to-speech conversion lies in isolating linguistic content from the input speech. Research on speech content disentanglement generally falls into a few main categories: (1) Information bottleneck: one line of work introduces a quantization layer \cite{softvc, hung-yi-vc, convspeak} or a low-dimensional hidden representation \cite{pingan-vc, sparc, ddsp}, often combined with proxy tasks such as ASR \cite{emovox, cosyvoice, cosyvoice2, cosyvoice3} or autoencoding \cite{autovc, triple-bottleneck, vevo}, to encourage the separation of content from style; (2) Signal perturbation: another family of methods \cite{nansy, nansy++, contentvec, seedvc} modifies style-related cues in the signal (e.g., pitch randomization, formant shifting) while preserving linguistic content, and trains the model to produce similar representations for the original and perturbed inputs; (3) Training loss design: alternative approaches incorporate objectives such as GAN loss \cite{stargan-vc, stargan-vc2, vaw-gan}, mutual information loss \cite{vqmivc, Yu-disentanglement, ddsvae}, or gradient reversal loss \cite{naturalspeech3, basetts} to explicitly promote disentanglement. However, content features extracted by these methods often still leak accent and emotion information, which degrades the quality of voice style conversion.

\subsection{Real-Time Voice Conversion}
The field of real-time voice conversion has seen rapid progress in both quality and latency \cite{streamvoice, streamvc, rt-vc, dualvc3}. For example, RT-VC \cite{rt-vc} achieves state-of-the-art real-time zero-shot voice conversion quality with a CPU latency of only 61.4ms. However, no existing work has addressed the problem of real-time zero-shot voice style conversion, where the goal is to modify not only timbre but also accent and emotion. This gap motivates our work, where we present StyleStream, the first system for real-time zero-shot voice style conversion.

\section{Method}
\subsection{Destylizer: Content-Style Disentanglement}
\begin{figure*}[t]
    \centering
    \includegraphics[scale=0.6]{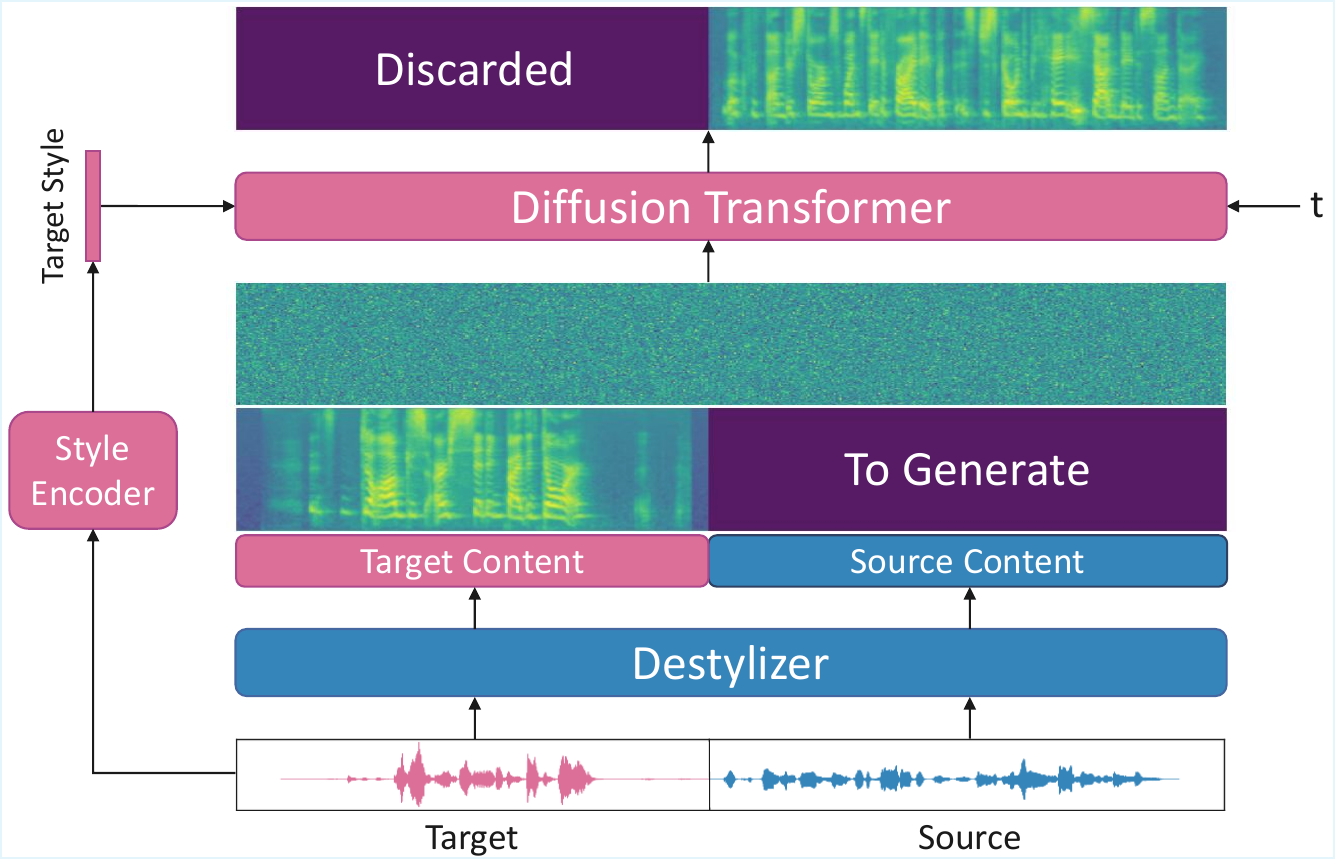}
    \caption{Stylizer architecture. The Stylizer contains a style encoder and a diffusion transformer.}
    \label{Stylizer}
\end{figure*}
To achieve clean content-style disentanglement, we introduce the Destylizer (Figure \ref{Destylizer}). It consists of a frozen HuBERT-Large encoder \cite{hubert} followed by several conformer blocks \cite{conformer}. Similar to CosyVoice 2 \cite{cosyvoice2}, the Destylizer and the FSQ module together form the ASR encoder. The continuous output features of the Destylizer, denoted as content features $f_c$, are first projected down to a $D$-dimensional space, and each dimension is quantized to the nearest integer in the range $[-K_i, K_i]$, where 
\begin{equation}
    V_i=2K_i+1, \quad K_i\in \mathbb{Z}_{>0}
\end{equation}
denotes the number of codes for the $i$-th dimension. The quantized low-dimensional features are then projected back to the original dimension and passed to the ASR decoder to predict text character tokens. The whole pipeline is trained end-to-end with a sequence-to-sequence ASR loss. At inference time, we use only the Destylizer to extract disentangled content features, operating at a sampling rate of 50 Hz.

There are three keys to successful content-style disentanglement: 

\textbf{Text Supervision.} Unlike Vevo \cite{vevo}, whose content tokenizer is trained in a purely self-supervised manner, we follow CosyVoice 2 and impose text supervision for training the Destylizer. This directs linguistic content through the FSQ bottleneck, while style is suppressed since it does not aid ASR prediction and the bottleneck provides limited capacity. In contrast, Vevo relies on a reconstruction objective with no explicit cue for separating content from style, causing style leakage and partial loss of linguistic content, which degrades disentanglement quality, as shown in Section \ref{exp}. 

\textbf{Compact Codebook.} Unlike CosyVoice 2, where a large codebook size (6561) is used, we use a much smaller codebook to enforce a narrower information bottleneck, following the intuition of AutoVC \cite{autovc}. Specifically, the vocabulary size $V$ for FSQ levels $[V_1, V_2, ..., V_D]$ is given by:
\begin{equation}
    V = \prod_{i=1}^DV_i
\end{equation}
Adjusting $D$ and $V_i$ controls the bottleneck width. As demonstrated in Section \ref{sec:content_feature_ablations}, CosyVoice 2's ``semantic tokens" remain largely entangled with style information, whereas our Destylizer provides much cleaner content features due to a compact codebook.

\textbf{Continuous Pre-quantization Features.} Unlike Vevo \cite{vevo} or CosyVoice 2 \cite{cosyvoice2}, which use discrete tokens as content, we instead adopt the continuous representations immediately before FSQ, inspired by SoftVC \cite{softvc}. SoftVC demonstrated that such “soft units” strike an effective balance between disentanglement and content preservation. Moreover, as shown in Section \ref{sec:content_feature_ablations}, our continuous features contain even less style information than the discrete tokens of Vevo and CosyVoice 2, while using FSQ indices directly as content leads to unintelligible speech.

\subsection{Stylizer: Stylized Acoustic Modeling}

After obtaining disentangled content features, the Stylizer generates mel-spectrograms conditioned on the target style. As shown in Figure \ref{Stylizer}, it consists of two components:

\textbf{Diffusion Transformer.}
We use a diffusion transformer (DiT) \cite{dit} as the backbone of the Stylizer, as it has demonstrated strong performance in in-context voice style cloning \cite{voicebox, e2tts, f5tts}. The Stylizer is trained with a spectrogram in-painting objective: given a temporal binary mask $m$, the model reconstructs the masked segment $m\odot x_1$ of a mel-spectrogram $x_1\in\mathbb{R}^{F\times T}$, conditioned on the unmasked context $(1-m)\odot x_1$, the content features $f_c\in\mathbb{R}^{D_c\times T}$, and a style embedding $e$ (described below), where $\odot$ denotes elementwise multiplication. 

Concretely, the noisy mel-spectrogram $x_t$, context $(1-m)\odot x_1$ and content features $f_c$ are concatenated along the channel dimension to form the DiT input. To ensure compatibility, the mel-spectrogram is calculated at the same rate as $f_c$ (50 Hz). The flow time step $t\sim \mathcal{U}[0,1]$ is embedded with a sinusoidal positional encoding, and added to the style embedding, which is then integrated into DiT via adaLN-Zero \cite{dit}. 

During training, we adopt the conditional flow matching (CFM) loss with an Optimal Transport (OT) path formulation. Let $x_1 \sim q(x)$ denote the ground-truth mel-spectrogram drawn from the data distribution, and $x_0 \sim p(x)=\mathcal{N}(0, I)$ be the standard normal prior. We sample a time step $t \sim \mathcal{U}[0, 1]$ and define the OT flow path as $\psi_t(x) = (1-t)x + tx_1$. Consequently, the noisy mel-spectrogram state at time $t$ is generated as $x_t = \psi_t(x_0)$. The predicted vector field is defined as:
\begin{equation}
\hat v
=
v_\theta\bigl(
\psi_t(x_0), t;\,
f_c,\,
(1-m)\odot x_1,\,
e
\bigr) 
\end{equation}
where $m$ is the binary mask indicating the generation target, $f_c$ represents the content features, and $e$ denotes the style embedding. The training objective is to minimize the difference between $\hat v$ and the target velocity, computed strictly on the masked regions:

\begin{equation}
\mathcal{L}_{\mathrm{CFM}}(\theta)
=
\mathbb{E}_{t, x_1, x_0, m}
\Bigl\lVert
m \odot
\bigl(
\hat v
-
\frac{d}{dt}\psi_t(x_0)
\bigr)
\Bigr\rVert^2 
\end{equation}

During inference, we concatenate the content features of the target and source utterances along the time axis. The target utterance provides the mel-spectrogram context and the style embedding, while the source region is masked and generated by the Stylizer. To balance diversity with fidelity, we employ Classifier-Free Guidance (CFG):
\begin{equation}
    v_{\theta, \text{CFG}}=v_{\theta}(x_t, t; c) + \alpha(v_{\theta}(x_t, t; c) - v_{\theta}(x_t, t; \varnothing))
\end{equation}
where $\varnothing$ denotes null condition, $c$ denotes all conditions, including $f_c, (1-m)\odot x_1, e$, and $\alpha$ is the CFG strength.

\textbf{Style Encoder.}
For style conditioning, we introduce a style encoder to capture global style attributes. Its architecture follows WavLM-TDNN\footnote{\url{https://huggingface.co/microsoft/wavlm-base-plus-sv}}: representations from frozen WavLM layers \cite{wavlm} are aggregated using learnable coefficients, and the aggregated features are then passed to a Time Delay Neural Network (TDNN) \cite{tdnn}. Attentive statistics pooling \cite{atsp} is applied to obtain the final style embedding, which is fused into DiT through adaLN-zero. The style encoder is trained jointly with the DiT in an end-to-end manner. 

\subsection{Vocoder}
We trained a causal vocoder to synthesize 16 kHz speech from 50 Hz mel-spectrograms for real-time inference. The design follows Vocos \cite{vocos}, with all convolution layers replaced by causal convolution. The training objective combines GAN loss, reconstruction loss, and feature matching loss, using the same configurations as Vocos. The original Vocos checkpoint\footnote{\url{https://github.com/gemelo-ai/vocos}} is used as a warm start.

\subsection{Real-Time Design}
To enable streaming, we apply a chunked-causal attention mask to both the Destylizer and Stylizer, where each chunk attends to its own features and all preceding chunks, but not to future chunks. In the Destylizer, the HuBERT layers are unfrozen and made chunked-causal, and all convolution layers are converted to causal. The streaming Destylizer is initialized from its non-streaming checkpoint and trained with a mean squared error (MSE) distillation loss against the non-streaming teacher’s content features. Once trained, the streaming Stylizer, warm-started from its non-streaming counterpart, is trained on top of the streaming Destylizer’s outputs. 

For real-time inference, we maintain a fixed-length target utterance and a ring buffer of input chunks of content features. This provides sufficient past context for the Stylizer while keeping latency manageable. The end-to-end latency $L$ is calculated as:
\begin{equation}
    L = t_{\text{chunksize}}+t_{\text{proc}}
    \label{latency}
\end{equation}
where $t_{\text{chunksize}}$ is the input speech chunk size and $t_{\text{proc}}$ is the processing time per chunk. As long as $t_{\text{proc}}<t_{\text{chunksize}}$, the model is streamable. 

\section{Experimental Setup}
\begin{table*}[t]
\centering
\small
\caption{Results for zero-shot voice style conversion. The best results are shown in \textbf{bold}, and the second best results are \underline{underlined}.}
\label{tab:model_comparison}
\resizebox{\textwidth}{!}{
\begin{tabular}{lcccccccc}
\toprule
\textbf{Model} & \textbf{WER ($\%$)} $\downarrow$ & \textbf{S-SIM} $\uparrow$ & \textbf{A-SIM} $\uparrow$ & \textbf{E-SIM} $\uparrow$ & \textbf{NMOS} $\uparrow$ & \textbf{A-SMOS} $\uparrow$ & \textbf{E-SMOS} $\uparrow$ & \textbf{S-SMOS} $\uparrow$ \\
\midrule
Ground Truth & 3.8 & --- & --- & --- & 3.71{\scriptsize\textcolor{gray}{$\pm$.11}} & --- & --- & --- \\
\midrule
FACodec \cite{naturalspeech3} & 15.5 & 0.763 & 0.408 & 0.668 & 2.55{\scriptsize\textcolor{gray}{$\pm$.13}} & 2.02{\scriptsize\textcolor{gray}{$\pm$.13}} & 2.76{\scriptsize\textcolor{gray}{$\pm$.14}} & 2.98{\scriptsize\textcolor{gray}{$\pm$.14}} \\
CosyVoice 2.0 \cite{cosyvoice2} & \underline{9.5} & 0.794 & 0.450 & 0.655 & \textbf{3.47}{\scriptsize\textcolor{gray}{$\pm$.11}} & 2.26{\scriptsize\textcolor{gray}{$\pm$.13}} & 2.58{\scriptsize\textcolor{gray}{$\pm$.15}} & 3.25{\scriptsize\textcolor{gray}{$\pm$.14}} \\
SeedVC v2\footref{seedvc} & 21.7 & 0.766 & 0.549 & 0.688 & 2.65{\scriptsize\textcolor{gray}{$\pm$.12}} & 3.34{\scriptsize\textcolor{gray}{$\pm$.12}} & 3.21{\scriptsize\textcolor{gray}{$\pm$.13}} & 3.11{\scriptsize\textcolor{gray}{$\pm$.13}} \\
Vevo \cite{vevo} & 17.5 & 0.818 & 0.596 & 0.712 & 3.38{\scriptsize\textcolor{gray}{$\pm$.12}} & 3.93{\scriptsize\textcolor{gray}{$\pm$.11}} & 3.49{\scriptsize\textcolor{gray}{$\pm$.13}} & 3.76{\scriptsize\textcolor{gray}{$\pm$.13}} \\
Vevo 1.5 \cite{amphion2} & 20.5 & 0.798 & 0.554 & 0.683 & 3.10{\scriptsize\textcolor{gray}{$\pm$.12}} & 3.15{\scriptsize\textcolor{gray}{$\pm$.14}} & 3.15{\scriptsize\textcolor{gray}{$\pm$.14}} & 3.34{\scriptsize\textcolor{gray}{$\pm$.14}} \\
\midrule
StyleStream (streaming) & 15.3 & \textbf{0.855} & \underline{0.635} & \underline{0.803} & 3.34{\scriptsize\textcolor{gray}{$\pm$.13}} & \underline{4.28}{\scriptsize\textcolor{gray}{$\pm$.09}} & \underline{4.37}{\scriptsize\textcolor{gray}{$\pm$.09}} & \underline{4.29}{\scriptsize\textcolor{gray}{$\pm$.10}} \\
StyleStream (offline) & \textbf{9.2} & \underline{0.852} & \textbf{0.640} & \textbf{0.827} & \underline{3.42}{\scriptsize\textcolor{gray}{$\pm$.12}} & \textbf{4.32}{\scriptsize\textcolor{gray}{$\pm$.09}} & \textbf{4.42}{\scriptsize\textcolor{gray}{$\pm$.09}} & \textbf{4.36}{\scriptsize\textcolor{gray}{$\pm$.10}} \\

\bottomrule
\end{tabular}}
\end{table*}
\subsection{Dataset}\label{sec:data}
For the training data, the Stylizer is trained using the English portion of Emilia \cite{emilia}, which contains 50k hours of diverse in-the-wild English speech. The Destylizer is trained on a combined training set from LibriTTS \cite{libritts}, MSP-Podcast \cite{msp-podcast}, and GLOBE \cite{globe}, totaling approximately 1300 hours of speech that covers a wide range of speakers, emotions and accents. We refer to this combined dataset as LMG. LMG is also used for training in the analysis and ablation studies (Sections \ref{sec:content_feature_ablations}, \ref{ablation}, \ref{destylizer_data}). 

For evaluation, we randomly sample 300 source utterances from a combination of the Emotion Speech Dataset (ESD) \cite{esd}, GLOBE-test, and LibriTTS-test-clean. We also select 10 target utterances from ESD, RAVDESS \cite{ravdess}, GLOBE-test, and L2-ARCTIC \cite{l2-arctic}, covering 5 emotions (happy, angry, sad, fearful, calm) and 5 accents (British, American, Indian, Arabic, Mandarin). This results in $300\times10=3000$ source-target pairs, which we denote as StyleStream-Test. All the objective and subjective evaluations in Section \ref{main} are conducted on StyleStream-Test. Moreover, to evaluate content-style disentanglement (Section \ref{sec:content_feature_ablations}), we train style classifiers on top of the content features extracted from different models. The speaker classifier is trained on the VoxCeleb training set \cite{voxceleb} and evaluated on its test set, which contains 1,251 speakers. The accent classifier is trained on L2-ARCTIC, which includes 6 accents; 6 speakers (one per accent) are held out for testing, and the remaining 18 speakers are used for training. The emotion classifier is trained on EmoV-DB \cite{emov-db}, which covers 5 emotions; 2 speakers (each with all 5 emotions) are held out for testing, and the remaining 18 speakers are used for training.

\subsection{Training}\label{training}

\begin{table*}[t]
\centering
\small
\caption{Chunk size vs. quality tradeoff}
\label{tab:chunksize-quality-tradeoff}

\setlength{\tabcolsep}{10pt} 
\begin{tabular}{cccccc}
\toprule
\textbf{Chunk size (ms)} & \textbf{WER ($\%$)} $\downarrow$ & \textbf{S-SIM} $\uparrow$ & \textbf{A-SIM} $\uparrow$ & \textbf{E-SIM} $\uparrow$ & \textbf{UTMOS} $\uparrow$ \\
\midrule
200 & 19.8 & 0.797 & 0.586 & 0.704 & 2.33{\scriptsize\textcolor{gray}{$\pm$.02}} \\
400 & 15.0 & 0.806 & 0.597 & 0.720 & 2.69{\scriptsize\textcolor{gray}{$\pm$.03}} \\
600 & 13.6 & 0.811 & 0.602 & 0.734 & 2.85{\scriptsize\textcolor{gray}{$\pm$.03}} \\
800 & 12.7 & 0.816 & 0.607 & 0.745 & 2.94{\scriptsize\textcolor{gray}{$\pm$.03}} \\
1000 & 12.6 & 0.820 & 0.613 & 0.751 & 3.00{\scriptsize\textcolor{gray}{$\pm$.03}} \\
\bottomrule
\end{tabular}
\end{table*}

\textbf{Destylizer.} We take the 18th layer of HuBERT-Large-ASR\footnote{\url{https://huggingface.co/facebook/hubert-large-ls960-ft}}
 as input to six conformer blocks, followed by an FSQ module and four transformer decoder layers (hidden size 768, FFN size 3072). ALiBi positional encoding \cite{alibi} is adopted to reduce long-context reliance, making the model suitable for length extrapolation and real-time inference. FSQ levels are set to [5,3,3], yielding a compact codebook of 45 codes. The model is trained for 100k steps on 8 NVIDIA RTX A6000 GPUs (batch size 32) using AdamW \cite{adamw} with a peak learning rate of 1e-4, 4k warm-up steps, and cosine annealing. The streaming variant uses a chunk size of 600 ms but otherwise follows the same training configuration.

\textbf{Stylizer.} The Stylizer consists of 16 transformer layers (hidden size 768, FFN size 3072). Speech is sampled at 16 kHz and converted to 100-bin mel-spectrograms with a hop size of 320, producing a 50 Hz frame rate. During training, $70-100\%$ of mel-spectrogram frames are randomly masked for inpainting. To support CFG inference, content features are dropped with probability 0.2, while context spectrograms and style embeddings are dropped with probability 0.3. Training uses 6s segments for 400k steps on 8 A6000 GPUs (batch size 64), with AdamW at a peak learning rate of 1e-4, 2k warm-up steps, and cosine annealing. The streaming Stylizer also uses a 600 ms chunk size under the same configuration. Throughout the experiments, the CFG strength is fixed at 2 and the Number of Function Evaluations (NFE) is set to 16, utilizing the standard Euler sampling method.

\textbf{Vocoder.} 
We follow the design choice of Vocos \cite{vocos}, but modify the convolution layers in ConvNext \cite{convnext} blocks into causal convolutions. Training is initialized from the official checkpoint and performed on the LibriTTS training set for 100k steps, using 2 A6000 GPUs with a batch size of 64 and 2s input segments. The mel-spectrogram setup matches that of the Stylizer, and all remaining training configurations follow Vocos.

\section{Baselines}\label{baselines}
We compare StyleStream against several representative voice conversion and voice style conversion systems. \textbf{FACodec}\footnote{\url{https://github.com/open-mmlab/Amphion/tree/main/models/codec/ns3_codec}} \cite{naturalspeech3} disentangles speech into phoneme content, normalized pitch, and speaker residuals using an information bottleneck with a gradient reversal layer, and performs conversion by swapping speaker embeddings. \textbf{CosyVoice 2.0}\footnote{\url{https://github.com/FunAudioLLM/CosyVoice}} \cite{cosyvoice2} employs a supervised semantic speech tokenizer trained with an ASR loss and finite scalar quantization (FSQ) with a 6561-entry codebook, followed by a flow-matching transformer trained with a spectrogram in-painting objective conditioned on semantic tokens and speaker embeddings. \textbf{SeedVC v2}\footnote{\label{seedvc}\url{https://github.com/Plachtaa/seed-vc}} is an upgraded version of SeedVC \cite{seedvc} that supports voice style conversion, for which no accompanying paper has been released. \textbf{Vevo}\footnote{\url{https://github.com/open-mmlab/Amphion/tree/main/models/vc/vevo}} \cite{vevo}, the previous state-of-the-art voice style conversion system, leverages self-supervised content tokens with a 32-code codebook, autoregressive content–style modeling, and Voicebox-style \cite{voicebox} acoustic modeling. Finally, \textbf{Vevo 1.5}\footnote{\url{https://github.com/open-mmlab/Amphion/tree/main/models/svc/vevosing}} \cite{amphion2} extends Vevo to both voice and singing voice conversion by introducing a prosody tokenizer that captures coarse-grained melody contours.

\subsection{Metrics}
\begin{table}[t]
\centering
\small
\caption{Chunk size vs. processing time on RTX 4060 and RTX A6000. Entries show mean $\pm$ standard deviation over runs.}
\label{tab:chunksize_processing_time}
\begin{tabular}{ccc}
\toprule
\textbf{Chunk size (ms)} & \textbf{RTX 4060} (s) & \textbf{RTX A6000} (s) \\
\midrule
100 & 0.537{\scriptsize\textcolor{gray}{$\pm$.053}} & 0.427{\scriptsize\textcolor{gray}{$\pm$.013}} \\
200 & 0.574{\scriptsize\textcolor{gray}{$\pm$.059}} & 0.429{\scriptsize\textcolor{gray}{$\pm$.009}} \\
300 & 0.615{\scriptsize\textcolor{gray}{$\pm$.057}} & 0.432{\scriptsize\textcolor{gray}{$\pm$.010}} \\
400 & 0.647{\scriptsize\textcolor{gray}{$\pm$.090}} & 0.441{\scriptsize\textcolor{gray}{$\pm$.008}} \\
500 & 0.653{\scriptsize\textcolor{gray}{$\pm$.051}} & 0.431{\scriptsize\textcolor{gray}{$\pm$.009}} \\
600 & 0.668{\scriptsize\textcolor{gray}{$\pm$.048}} & 0.429{\scriptsize\textcolor{gray}{$\pm$.006}} \\
700 & 0.673{\scriptsize\textcolor{gray}{$\pm$.041}} & 0.429{\scriptsize\textcolor{gray}{$\pm$.002}} \\
800 & 0.673{\scriptsize\textcolor{gray}{$\pm$.040}} & 0.431{\scriptsize\textcolor{gray}{$\pm$.004}} \\
\bottomrule
\end{tabular}
\end{table}
For objective evaluation, we report word error rate (WER) as a measure of intelligibility, computed with Whisper-large-v3 \cite{whisper}. To evaluate style similarity, we extract embeddings for speaker\footnote{\url{https://github.com/resemble-ai/Resemblyzer}}, accent\footnote{\url{https://huggingface.co/Jzuluaga/accent-id-commonaccentecapa}} \cite{commonaccent}, and emotion\footnote{\url{https://github.com/ddlBoJack/emotion2vec}} \cite{emotion2vec}, and compute cosine similarity between generated and target speech, yielding speaker similarity (S-SIM), accent similarity (A-SIM), and emotion similarity (E-SIM). For subjective evaluation, we conduct Mean Opinion Score (MOS) tests on a 5-point scale, including naturalness MOS (N-MOS) for converted speech and similarity MOS (SMOS) for speaker (S-SMOS), accent (A-SMOS), and emotion (E-SMOS) relative to the target. All subjective evaluations are carried out with crowd-sourced listeners recruited via Prolific\footnote{\url{https://www.prolific.com/}}, where all participants are based in the US or UK, are native English speakers, and are familiar with common English accents. For each test condition, we collect a total of 400 ratings per model. In N-MOS, listeners rate the overall naturalness of each generated sample from 1 (completely unnatural) to 5 (completely natural). In SMOS, listeners rate the similarity between a converted sample and the corresponding target on the same 5-point scale; for accent similarity, listeners are instructed to ignore speaker timbre, emotion, and recording quality, and focus solely on accent similarity.

\section{Results}\label{exp}

\begin{table*}[t]
\centering
\small
\caption{Stylizer performance trained on different content features.}
\label{tab:ablation_results}
\setlength{\tabcolsep}{12pt} 
\begin{tabular}{lccccc}
\toprule
\textbf{Content Features} & \textbf{WER ($\%$)} $\downarrow$ & \textbf{S-SIM} $\uparrow$ & \textbf{A-SIM} $\uparrow$ & \textbf{E-SIM} $\uparrow$ & \textbf{UTMOS} $\uparrow$ \\
\midrule
HuBERT-Large-ASR 18th & 12.5 & 0.731 & 0.389 & 0.627 & \textbf{3.28}{\scriptsize\textcolor{gray}{$\pm$.020}} \\
Vevo Continuous & 12.6 & 0.767 & 0.420 & 0.660 & 2.92{\scriptsize\textcolor{gray}{$\pm$.029}} \\
Vevo Indices & 25.6 & 0.757 & 0.564 & 0.657 & 2.65{\scriptsize\textcolor{gray}{$\pm$.036}} \\
Destylizer (offline) & \textbf{10.7} & \textbf{0.837} & \textbf{0.626} & \textbf{0.733} & 3.22{\scriptsize\textcolor{gray}{$\pm$.029}} \\

\bottomrule
\end{tabular}
\end{table*}

\begin{table*}[t]
\centering
\small
\caption{Style classification accuracy across different content features. ``Acc." stands for accuracy.}
\label{tab:model_performance}
\setlength{\tabcolsep}{12pt} 
\begin{tabular}{lccc}
\toprule
\textbf{Content Features} & \textbf{Accent Acc.} $\downarrow$ & \textbf{Emotion Acc.} $\downarrow$ & \textbf{Speaker Acc.} $\downarrow$ \\
\midrule
HuBERT-Large-ASR 18th & 78.00\% & 85.60\% & 86.60\% \\
Vevo Continuous & 68.93\% & 78.73\% & 64.76\% \\
Vevo Indices & 55.60\% & 53.40\% & 23.00\% \\
CosyVoice 2.0 Indices & 50.60\% & 56.20\% & 25.00\% \\
Destylizer (offline) & 43.50\% & 47.60\% & 3.50\% \\
Destylizer (streaming) & \textbf{33.80\%} & \textbf{37.90\%} & \textbf{3.01\%} \\
\bottomrule
\end{tabular}
\end{table*}

\subsection{Zero-Shot Voice Style Conversion}
\label{main}

The main results are shown in Table~\ref{tab:model_comparison}. Note that StyleStream does not provide independent control over individual style factors; all evaluations reflect holistic target-style cloning, where speaker identity, accent, and emotion are transferred jointly. Overall, StyleStream consistently outperforms prior methods across intelligibility and multiple style similarity metrics, demonstrating strong zero-shot generalization in both offline and streaming settings.

In terms of intelligibility, the offline StyleStream achieves the lowest WER ($9.2\%$), substantially outperforming the previous state-of-the-art Vevo. Despite operating in a chunk-causal manner, the streaming variant remains competitive with Vevo, highlighting the effectiveness of the Destylizer in preserving linguistic content under streaming constraints. In contrast, methods such as SeedVC and FACodec exhibit higher WER, indicating weaker content preservation.

For overall style fidelity, StyleStream achieves the highest speaker, accent, and emotion similarity scores, as measured by both objective and subjective metrics. Specifically, the offline variant attains the best S-SIM, A-SIM, and E-SIM, while both offline and streaming models substantially outperform all baselines in S-SMOS, A-SMOS, and E-SMOS. These improvements are consistently reflected in both objective and subjective evaluations, indicating higher-fidelity and more perceptually accurate reproduction of the target speaking style.

Taken together, these results show that StyleStream achieves a more favorable balance between content preservation and holistic style fidelity than existing zero-shot voice style conversion systems, while maintaining robust performance in real-time streaming scenarios.

\subsection{Streaming Analysis}

\subsubsection{Latency} We measure the end-to-end streaming latency on a single NVIDIA RTX A6000 GPU. Using NFE=16, a chunk size of $t_{\text{chunksize}}=600$ms, a 5s target segment, and a 5s content ring buffer, with the target style embedding pre-extracted for inference, the average processing time per chunk is $t_{\text{proc}}=412.7$ms. Since $t_{\text{proc}}<t_{\text{chunksize}}$, streaming is feasible. By Equation~\ref{latency}, the resulting end-to-end latency is $L=600+412.7=1012.7$ms.

\subsubsection{Chunksize-Latency Analysis} We benchmark the processing time ($t_{\text{proc}}$) across varying chunk sizes on both a server-grade NVIDIA A6000 and a consumer-grade RTX 4060 Laptop GPU (NFE fixed at 16). As shown in Table \ref{tab:chunksize_processing_time}, $t_{\text{proc}}$ on the A6000 remains effectively constant ($\approx 0.43$s) when varying the chunk size from 100ms to 800ms. This indicates that the DiT inference is dominated by fixed kernel launch overheads rather than computational saturation on high-end hardware. In contrast, the RTX 4060 exhibits a linear increase in processing time with chunk size, reflecting a compute-bound regime characteristic of consumer-grade deployment.

\subsubsection{Chunksize-Quality Trade-off} To decouple the effects of context length from training-inference mismatch, we analyze the chunksize-quality trade-off by simulating streaming inference using the offline checkpoint across chunk sizes ranging from 200ms to 1000ms (Table \ref{tab:chunksize-quality-tradeoff}). We observe a monotonic improvement in both intelligibility (WER) and style similarity as chunk size increases. This confirms that extended temporal context is essential for capturing accent and emotion and minimizing discontinuities at chunk boundaries.

\subsection{Content-Style Disentanglement Analysis}\label{sec:content_feature_ablations}

\begin{table*}[t]
\centering
\small
\caption{Ablation studies on the use of style encoder and the choice of Destylizer content features.}
\label{tab:ablation_stylizer_results}
\setlength{\tabcolsep}{12pt} 
\begin{tabular}{lccccc}
\toprule
\textbf{Model} & \textbf{WER ($\%$)} $\downarrow$ & \textbf{S-SIM} $\uparrow$ & \textbf{A-SIM} $\uparrow$ & \textbf{E-SIM} $\uparrow$ & \textbf{UTMOS} $\uparrow$ \\
\midrule
StyleStream (offline) & \textbf{10.7} & \textbf{0.837} & \textbf{0.626} & \textbf{0.733} & 3.22{\scriptsize\textcolor{gray}{$\pm$.029}} \\
w/o Style Emb & 15.3 & 0.775 & 0.509 & 0.653 & \textbf{3.47}{\scriptsize\textcolor{gray}{$\pm$.024}} \\
w/ FSQ Indices & 123.5 & 0.829 & 0.573 & 0.717 & 2.41{\scriptsize\textcolor{gray}{$\pm$.024}} \\
\bottomrule
\end{tabular}
\end{table*}

\begin{table*}[t]
\centering
\small
\caption{Ablation on Destylizer FSQ bottleneck size. ``*" denotes the architecture used in all main experiments.}
\label{tab:fsq_ablation_results}
\setlength{\tabcolsep}{12pt} 
\begin{tabular}{lccccc}
\toprule
\textbf{FSQ Levels} & \textbf{WER ($\%$)} $\downarrow$ & \textbf{S-SIM} $\uparrow$ & \textbf{A-SIM} $\uparrow$ & \textbf{E-SIM} $\uparrow$ & \textbf{UTMOS} $\uparrow$ \\
\midrule
{[}7,5,5,5,5{]} & 13.5 & 0.745 & 0.439 & 0.638 & \textbf{3.58}{\scriptsize\textcolor{gray}{$\pm$.022}} \\
{[}5,5,3,3{]} & 14.9 & 0.822 & 0.600 & 0.725 & 3.17{\scriptsize\textcolor{gray}{$\pm$.031}} \\
{[}5,3,3{]}* & \textbf{10.7} & \textbf{0.837} & \textbf{0.626} & \textbf{0.733} & 3.22{\scriptsize\textcolor{gray}{$\pm$.029}} \\
{[}3,3,3{]} & 19.0 & 0.825 & 0.618 & 0.732 & 3.13{\scriptsize\textcolor{gray}{$\pm$.032}} \\
{[}5,3{]} & 101.4 & 0.834 & 0.582 & 0.714 & 2.70{\scriptsize\textcolor{gray}{$\pm$.029}} \\
\bottomrule
\end{tabular}
\end{table*}

In this section, we analyze how different speech content representations affect downstream voice style conversion performance, with a particular focus on their degree of content-style disentanglement. Since the Destylizer is explicitly designed to remove timbre, accent, and emotion information while preserving linguistic content, we evaluate disentanglement both indirectly (via Stylizer performance) and directly (via auxiliary style classification probes).

\subsubsection{Effect of Content Representations on Stylizer}

We first assess how different content features influence downstream Stylizer performance. To this end, we train the Stylizer from scratch on top of the Destylizer and several existing speech content representations, and evaluate WER, S-SIM, A-SIM, E-SIM, and UTMOS~\cite{saeki2022utmos}. Specifically, we compare against the continuous pre-quantization features and discrete representations from Vevo, as well as the raw 18th-layer features of HuBERT-Large-ASR, which are also incorporated as the first stage of the Destylizer.

All Stylizers are trained on the LMG dataset (Section~\ref{sec:data}) using identical training configurations (Section~\ref{training}), with training limited to 100k steps for fair comparison. Results are reported in Table~\ref{tab:ablation_results}.

Stylizers trained on Destylizer features achieve the best overall performance in WER, S-SIM, A-SIM, and E-SIM, while remaining competitive in UTMOS. In contrast, the HuBERT-based Stylizer yields substantially lower A-SIM and E-SIM, indicating limited style transfer capability. For Vevo, using discrete indices significantly improves A-SIM compared to continuous features, but this comes at the cost of degraded content preservation, as reflected by the highest WER among all methods.

\subsubsection{Disentanglement Hypothesis}

We hypothesize that the observed performance differences are closely linked to the degree of content-style disentanglement in the underlying representations. When disentanglement is poor, residual style information may leak into the content features and be exploited by the Stylizer during training, causing it to reconstruct speech with source-style attributes. As a result, inference fails to faithfully reflect the target style, leading to degraded style fidelity despite strong training performance.

To validate this hypothesis, we directly measure how much timbre, accent, and emotion information remain in each set of content features.

\subsubsection{Style Classification Probing}

We train an ECAPA-TDNN~\cite{tdnn} classifier on top of each content representation to predict speaker identity, accent, and emotion. Since these attributes are precisely what the Destylizer is designed to remove, lower classification accuracy indicates stronger disentanglement, with random-guess performance as the ideal target.

Results are shown in Table~\ref{tab:model_performance}. Both variants of the Destylizer consistently achieve the lowest classification accuracy across all three tasks. In particular, Destylizer features yield approximately $\sim$3\% accuracy on speaker classification, compared to 20\% or higher for HuBERT, Vevo, and CosyVoice~2.0, indicating that most speaker information has been filtered out.

The slightly above-random accuracy observed for the Destylizer may stem from residual correlations between prosody or duration and the extracted content features. This is expected, as the non-autoregressive, streaming-oriented design preserves the source utterance duration. Nevertheless, the Destylizer yields the cleanest content representations among all compared methods, consistent with its superior downstream Stylizer performance.

\subsection{Ablations on Architecture}\label{ablation}

In this section, we present architectural ablations of StyleStream, all trained on the LMG dataset described in Section \ref{sec:data}. Specifically, we first evaluate a variant without style embeddings and another that uses FSQ quantized indices instead of pre-quantization continuous features. Results are reported in Table \ref{tab:ablation_stylizer_results}. Removing the style embedding significantly reduces S-SIM, A-SIM and E-SIM, showing that relying solely on the unmasked context mel-spectrogram is insufficient for style modeling. The style encoder is therefore essential for faithful style conversion. Moreover, replacing continuous features with FSQ indices severely degrades intelligibility, as shown by the high WER. At first glance, this may seem surprising, since Vevo achieves reasonable results using discrete indices. However, as shown in Table \ref{tab:model_performance}, our Destylizer’s continuous features already carry even less accent, emotion, and speaker information than Vevo’s discrete indices. Further quantization strips away critical linguistic content, which explains the poor downstream performance.

We also ablate the effect of the FSQ bottleneck size by varying the FSQ levels during Destylizer training and retraining the Stylizer on top (Table \ref{tab:fsq_ablation_results}). Enlarging the codebook to $7\times5\times5\times5\times5=4375$ improves UTMOS, since richer information is kept in the Destylizer features, making it easier for the Stylizer to reconstruct speech during training. However, this comes at the cost of a sharp drop in A-SIM and E-SIM, indicating substantial style leakage. Conversely, reducing the bottleneck size to $3\times3\times3=27$ degrades the intelligibility, and a further reduction to $5\times3=15$ severely damages content preservation, indicating an overly restrictive bottleneck. These results suggest that our chosen setting of [5, 3, 3] provides a favorable trade-off: compact enough to filter out most of the style, yet wide enough to preserve linguistic content.

\subsection{Ablations on Destylizer Training Data}\label{destylizer_data}

We study the impact of training data scale for the Destylizer by replacing the original 1.3k-hour LMG training set (Section \ref{sec:data}) with the larger Emilia-EN dataset (50k hours), while retraining the Stylizer on top of the resulting Destylizer. The comparative results are reported in Table~\ref{tab:training_data}. Scaling the Destylizer training data to 50k hours does not yield a significant overall improvement. Although content preservation improves marginally, with WER decreasing from 10.7\% to 10.3\%, the style disentanglement metrics (S-SIM and A-SIM) exhibit slight degradation.

We attribute this saturation effect to two main factors. First, the Destylizer performs a discriminative extraction task, compressing speech into content representations under ASR supervision, which fundamentally differs from the Stylizer’s generative modeling objective. Unlike diffusion-based generative models that benefit substantially from large-scale data to capture long-tail acoustic variations, the Destylizer’s content-style separation task saturates once robust disentanglement is learned, leading to diminishing returns with additional data. Second, the original 1.3k-hour dataset is strategically curated to maximize style diversity rather than raw duration, combining GLOBE for accent variability, MSP-Podcast for emotional diversity, and LibriTTS for clean timbre coverage. This composition already spans the style manifold required for effective disentanglement under a constrained bottleneck. In contrast, the substantially larger Emilia-EN dataset does not meaningfully expand this manifold for the Destylizer’s objective. Therefore, these results suggest that the 1.3k-hour LMG dataset represents an efficient and sufficient operating point for training the Destylizer, balancing data scale and disentanglement performance.

\begin{table}[t]
\centering
\caption{Ablation on the Destylizer training data configurations.}
\label{tab:training_data}
\resizebox{\linewidth}{!}{
\begin{tabular}{lcccc}
\toprule
\textbf{Training data} & \textbf{WER (\%) $\downarrow$} & \textbf{S-SIM $\uparrow$} & \textbf{A-SIM $\uparrow$} & \textbf{E-SIM $\uparrow$} \\
\midrule
LMG (1.3k h) & 10.7 & 0.837 & 0.626 & 0.733 \\
Emilia (50k h)                          & 10.3 & 0.816 & 0.594 & 0.687 \\
\bottomrule
\end{tabular}}
\end{table}

\subsection{Ablations on Target Utterance Lengths}
\begin{table}[t]
\centering
\caption{Ablation on target utterance lengths.}
\label{tab:target_lengths}
\resizebox{\linewidth}{!}{
\begin{tabular}{lcccc}
\toprule
\textbf{Model} & \textbf{WER (\%) $\downarrow$} & \textbf{S-SIM $\uparrow$} & \textbf{A-SIM $\uparrow$} & \textbf{E-SIM $\uparrow$} \\
\midrule
Target 5s (offline)   & 9.2  & 0.852 & 0.640 & 0.827 \\
Target 5s (streaming) & 15.3 & 0.856 & 0.635 & 0.803 \\
Target 2s (offline)   & 12.1 & 0.824 & 0.594 & 0.747 \\
Target 2s (streaming) & 19.2 & 0.833 & 0.598 & 0.728 \\
\bottomrule
\end{tabular}}
\end{table}
We further evaluate StyleStream on the same test set using truncated 2-second target reference utterances. The comparison with the 5-second reference setting is summarized in Table~\ref{tab:target_lengths}. As expected, reducing the target utterance length leads to consistent performance degradation across all metrics. In particular, both similarity scores and content preservation worsen for the 2-second setting in both offline and streaming modes. This degradation is linguistically intuitive. Accent and emotion are global style attributes that require sufficient temporal context to be reliably estimated. A 2-second reference often provides limited phoneme coverage, making it difficult to characterize a speaker’s accent distribution, and offers insufficient prosodic variation to robustly capture emotional cues. Consequently, shorter reference utterances constrain the model’s ability to infer stable style representations, resulting in reduced conversion quality.

\section{Conclusion}

In this work, we presented StyleStream, the first streamable zero-shot voice style conversion system capable of modifying timbre, accent, and emotion in real time, with an end-to-end latency of approximately 1 second. StyleStream is built upon two core components: a Destylizer, which performs explicit content-style disentanglement using ASR supervision and a compact finite scalar quantization (FSQ) bottleneck, and a Stylizer, which leverages a diffusion transformer to reintroduce target style conditioned on reference speech. By operating on continuous pre-quantization features, the proposed framework enables cleaner content-style separation and avoids artifacts commonly introduced by discrete tokenization. As a result, StyleStream achieves state-of-the-art voice style conversion performance across both objective and subjective evaluations, while maintaining full streamability, making it suitable for real-time applications.

\section{Generative AI Use Disclosure}
We made limited use of AI during the preparation of this paper. In particular, LLMs were used for grammar checking, rephrasing for clarity, and improving the readability of drafts. In addition, generative AI was employed to generate the female and angry emojis used in Figure \ref{system}.

\bibliographystyle{IEEEtran}
\bibliography{mybib}

\end{document}